\documentclass[aip,jap,twocolumn,amsmath,amssymb,reprint]{revtex4}
\usepackage{graphicx,color}
\usepackage{amsmath}
\usepackage{natbib}
\usepackage{epsfig}

\begin{document}

\date\today

\draft
\title{Particle charge in PK-4 dc discharge from ground-based and microgravity experiments}

\author{T.  Antonova,$ ^{1}$ S. A. Khrapak,$ ^{1}$ M. Y. Pustylnik,$ ^{1}$  M. Rubin-Zuzic,$ ^{1}$  H. M. Thomas,$ ^{1}$  A. M. Lipaev,$ ^{2}$  A. D. Usachev,$ ^{2}$  V. I.  Molotkov,$ ^{2}$\footnote{Our very much admired and respected friend and colleague Vladimir Ivanovich Molotkov passed away unexpectedly on July 11, 2019, when this paper was finalized for submission. } M. H. Thoma$ ^{3}$}

\address{ $ ^{1}$Institut  f\"ur Materialphysik im Weltraum, Deutsches Zentrum  f\"ur Luft- und Raumfahrt e.V. (DLR),82234 Wessling, Germany \\ $ ^{2}$Joint Institute for High Temperatures, Russian Academy of Sciences, 125412 Moscow, Russia \\ $ ^{3}$I. Physikalisches Institut, Justus-Liebig Universit\"at, 35392 Gie$\beta$en, Germany }

\begin{abstract}
The charge of microparticles immersed in the dc discharge of PK-4 experimental facility has been estimated using  the particle velocities from experiments performed on Earth and under microgravity conditions on the International Space Station (ISS). The theoretical model used for these estimates is based on the balance of the forces acting on a single particle in the discharge.  The model takes into account the radial dependence of the discharge parameters and describes reasonably well the experimental measurements.  
\end{abstract}

\maketitle

\section {Introduction}

Micrometer sized particles immersed into a plasma  form strongly coupled assemblies, which are the subject of active investigations in the field of complex (dusty) plasmas. Being relatively easy to resolve at a single particle level, such systems can be successfully used for modeling  the kinetic properties in condensed matter physics.~\cite{fortov, morfill} In order to better control and explain the behavior of the particle assemblies, it is necessary to have knowledge about the basic mechanisms of the plasma-particle interaction, especially the charging process. Similar to the previous work~\cite{khrapak2013} we estimate the charge of the microparticles drifting in the direct current (dc) discharge of the complex plasma laboratory Plasmakristall-4 (PK-4). The facility was installed in the Columbus module of the International Space Station (ISS) in November 2014.~\cite{pustylnik} It is an experimental laboratory with the possibility to provide a range of various experiments in the dc as well as radio-frequency (rf) low temperature discharge including different manipulation techniques (e.g. laser manipulation, thermal and electric disturbances). 
Having an elongated shape of the plasma chamber, this set-up is particularly suitable for studying the flow-related phenomena in fluid complex plasmas on the kinetic level. Since the installation of PK-4 onboard the ISS several scientific campaigns were performed with sets of different dedicated experiments (e.g. shear flows,~\cite{weber} low frequency waves~\cite{surabi,yaroshenko2019}). 

We concentrate on the analysis of a series of experiments with  microparticles drifting in the plasma chamber tube. In a previously published work~\cite{khrapak2013} it was observed that the particle drift velocities in the dc discharge of PK-4 under gravity conditions are always higher than those measured during parabolic flights in microgravity at identical experimental parameters. The developed analytical model provided the explanation of this phenomenon based on one of the main plasma-particle interaction mechanism -- the charging process. It was concluded that the observed difference in the particle drift velocities comes from the variation of particle positions in the discharge: under microgravity conditions particles are levitating in the plasma bulk close to the discharge axis; in the  ground-based laboratory they are shifted in the vertical direction towards  the chamber walls due to the force of gravity. Particle charge somewhat increases from the tube axis towards the periphery and this explains the higher drift velocities observed in laboratory conditions. These results show the importance of determination of the local distribution of the discharge parameters as well as of the particle charge. 

In this paper we present the analysis of particle drift flows in the  PK-4 tube from a series of experiments made in ground-based laboratory and under microgravity conditions onboard the ISS. Using a theoretical model of particle charging which takes into account the radial dependence of the discharge parameters within the discharge tube, we compare experimentally measured particle velocities with those given by the model and estimate the microparticle charge.

\section{Experimental procedure} 

The experiments have been performed in PK-4 onboard the ISS as well as using the Scientific Reference Model 1 (SRM 1) of PK-4 built identical to the Flight Model in the ground based laboratory. The plasma is generated inside an U-shaped glass tube (horizontal part  has a length of 35 cm) with a radius of 1.5 cm by a dc discharge with  high voltage (HV) in the range of 0--3 kV and current of 0--3 mA applied to the active electrode. The discharge tube on the ground is in horiziontal configuration.  Complex plasma is formed by injecting spherical melamine-formaldehyde particles of different sizes into the discharge chamber.

\begin{figure}
	\center
	\includegraphics[width=7.5cm]{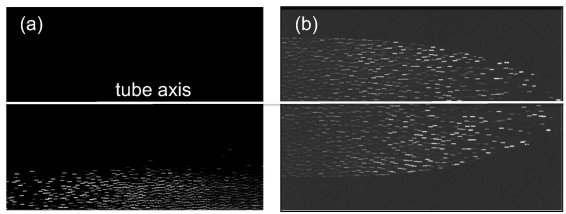}
	\caption{The examples of the drift flow of 3.4 $\mu$m diameter particles in argon dc discharge of PK-4 at 100 Pa pressure and 1 mA discharge current in  laboratory conditions (a) and under microgravity condition on ISS (b). The pictures are taken with 35 fps frame rate and 28 msec exposure time.  }
\end{figure}

Once particles are introduced into the plasma, they usually acquire a negative charge (due to highly mobile electrons) and drift from the cathode to anode with a typical velocity of 1 -- 10 cm/s. In laboratory conditions the microparticle cloud is shifted downwards (Fig.~1, a), because a significant radial electric field is required to compensate for  gravity. 
Under microgravity conditions the microparticles are located in the vicinity of the discharge axis (Fig.~1, b). Here, they form clouds of several layers with a distance between the microparticles (depending on the discharge parameters and microparticle size) ranging from 100 to 300 $\mu$m.  The microparticle suspensions are illuminated by a laser diode of 150 mW maximum power and the wavelength of 532 nm. The light scattered by the microparticles  is recorded by two video cameras, which allow to observe a field of view  (FoV)  of  $44\times 17$ mm$^{2}$ with a maximum resolution of  approximately 14 $\mu$m  and a frame rate of 35 fps. 

Besides dc discharge, PK-4  facility has the possibility to ignite an  inductively coupled  rf discharge with  81.36 MHz and maximal forward power of 5 W using two rf coils: one coil has a fixed position and the other can be moved along the discharge tube. A more detailed description of the PK-4 apparatus is presented in Ref.~\onlinecite{pustylnik}. 

One of the experiments performed on ISS was dedicated to estimate the velocities of the particles drifting in the dc discharge tube.  In order to have better statistics in velocity measurements and control over the microparticle suspension, a combined microparticle transport and  trapping technique (dc discharge transport and rf discharge trapping) were applied  (see Ref.~\onlinecite{pustylnik}, Section II H). 

In this method, the injected negatively charged microparticles were first transported to the rf discharge of the movable rf coil by the longitudinal electric field of the dc discharge. Then, the dc plasma  was switched off and microparticles were stored in the rf discharge. Afterwards a polarity-switched plasma  was ignited with 50$\%$ duty cycle and 500 Hz frequency and the  rf discharge was switched off. Next a dc discharge was  applied first with a negative current and the microparticles were drifting for a certain time in one direction. Later, the polarity was reversed and the microparticles were drifting in the opposite direction. This allowed us to get more statistics on microparticle drift velocity. 

The experimental procedure was similar to that  described previously.~\cite{surabi}  
But the experiments were performed in a broader range of parameters on board the ISS as well as in the ground laboratory. Neon and argon gases were chosen in the pressure range of 20 -- 100 Pa.  Monodisperse plastic (melamine formaldehyde)  microparticles  of three different sizes (1.3, 2.5 and 3.4 $\mu$m in diameter) were injected into a discharge.  
There was a small gas leak of about 0.1 sccm observed during the experiments as mentioned in Ref.~\onlinecite{surabi}.  The dc discharge was switched on with  different values of the discharge current (0.5--1.5 mA).  The duration time for the measurement procedure was between 1 to 3 seconds in each direction depending on the discharge current (the higher the discharge current,  the slower the particles are moving). The rf discharge was ignited with a constant power of 1 W for argon and 0.4~W for neon gas. The experimental procedure on the ISS  was similar to that performed in ground laboratory. On the ISS,  only particles with 3.4~$\mu$m in diameter were used for this experiment.  
  
\begin{figure}
	\center
	\includegraphics[width=7.5cm]{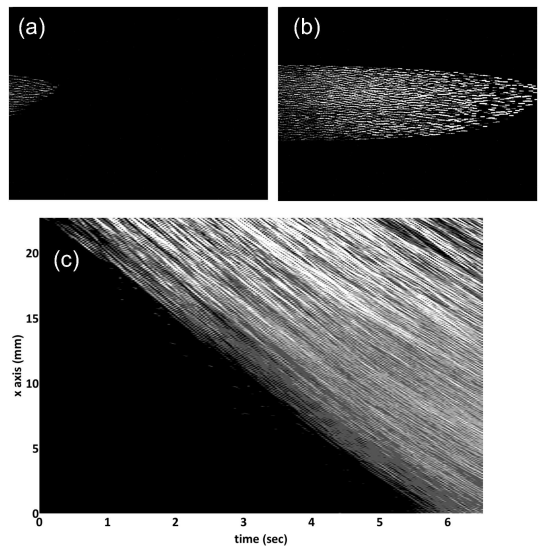}
	\caption{The upper pictures (a) and (b) show the front of the particle cloud entering and leaving the FoV (particles flow from left to right).   The lower picture (c) is the space-time diagram  for the sequence of frames (140 frames).  The pictures are taken from the experiment done on the ISS with the 3.4  $\mu$m diameter particles in neon dc discharge, 60 Pa pressure and 1 mA discharge current.  }
\end{figure}

Experimentally, the  particle velocities were estimated in two ways for comparison. In the first method the velocity was determined from the time of flight of the suspension through the FoV (Fig.~2 a,b).  Another method was relying on the slope in the space-time diagram of the microparticle motion. 
(Fig.~2c). If one has sequential images, where moving particles are seen of which the coordinates cannot be obtained (e.g. since they overlap), it is possible to gain additional dynamical information from the corresponding  ``space time diagrams''. 
The main idea is to measure the mean brightness in columns and rows as a function of time (image number). From the obtained images e.g. the mean velocity of the particle cloud can be measured from the slope seen in such a space time diagram.
 Here one needs to take into account  that the additional gas velocity due to the gas flow leak, as it has been mentioned in Ref.~\onlinecite{surabi}, was in the range of 0.6 to  1 cm/s depending on the gas pressure.

In order to characterize the discharge,  Langmuir probe measurements were performed in the particle-free plasma on the axis of the discharge tube at different experimental conditions  in a ground setup identical to the present PK-4 models.~\cite{fortovmorfill} The discharge parameters such as electron density, electron temperature and electric field values  measured in the pressure range of 20 -- 100 Pa in  neon and argon  gases at 1 mA discharge current are presented in Fig. 3 and Fig. 4,  respectively. 

\begin{figure}
	\center
	\includegraphics[width=7.5cm]{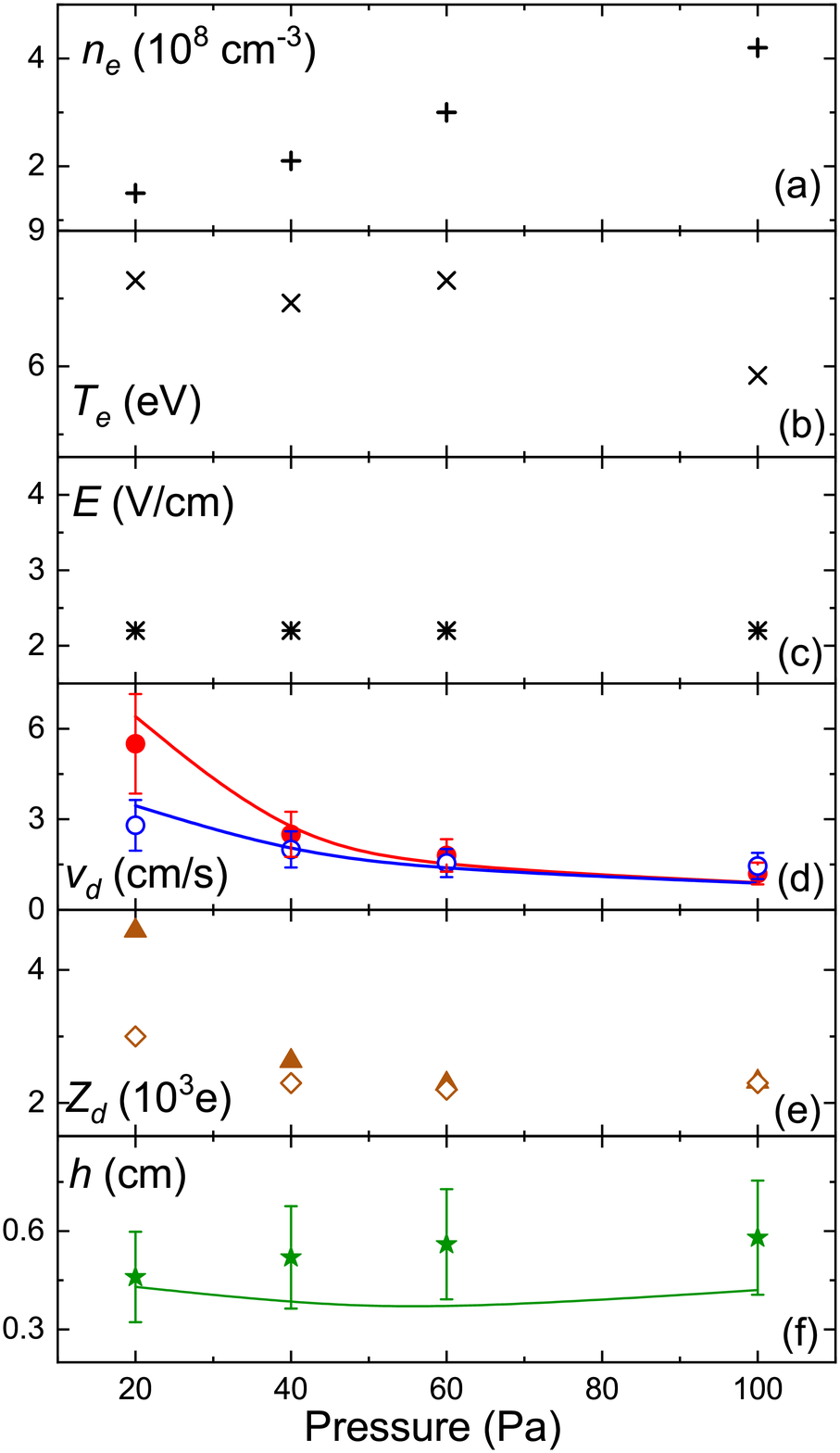}
	\caption{Pressure dependence on the main plasma parameters (electron density  $n_e$, electron temperature $T_e$ and electric filed $E$), experimentally measured (circles)  and theoretically estimated (curves) microparticle drift velocities $V_{d}$ for ground (red color) and microgravity (blue color) conditions with experimental uncertainties, theoretically estimated microparticle charge  $Z_{d}$ for ground (triangle) and microgravity (diamond) conditions, distance of the upper microparticle layer from the center of the tube $h$ measured on ground (stars) and theoretically estimated (curve). All values are presented for the case of neon gas, 1~mA discharge current  and 3.4 $\mu$m diameter microparticles. }
\end{figure}

\begin{figure}
	\center
	\includegraphics[width=7.5cm]{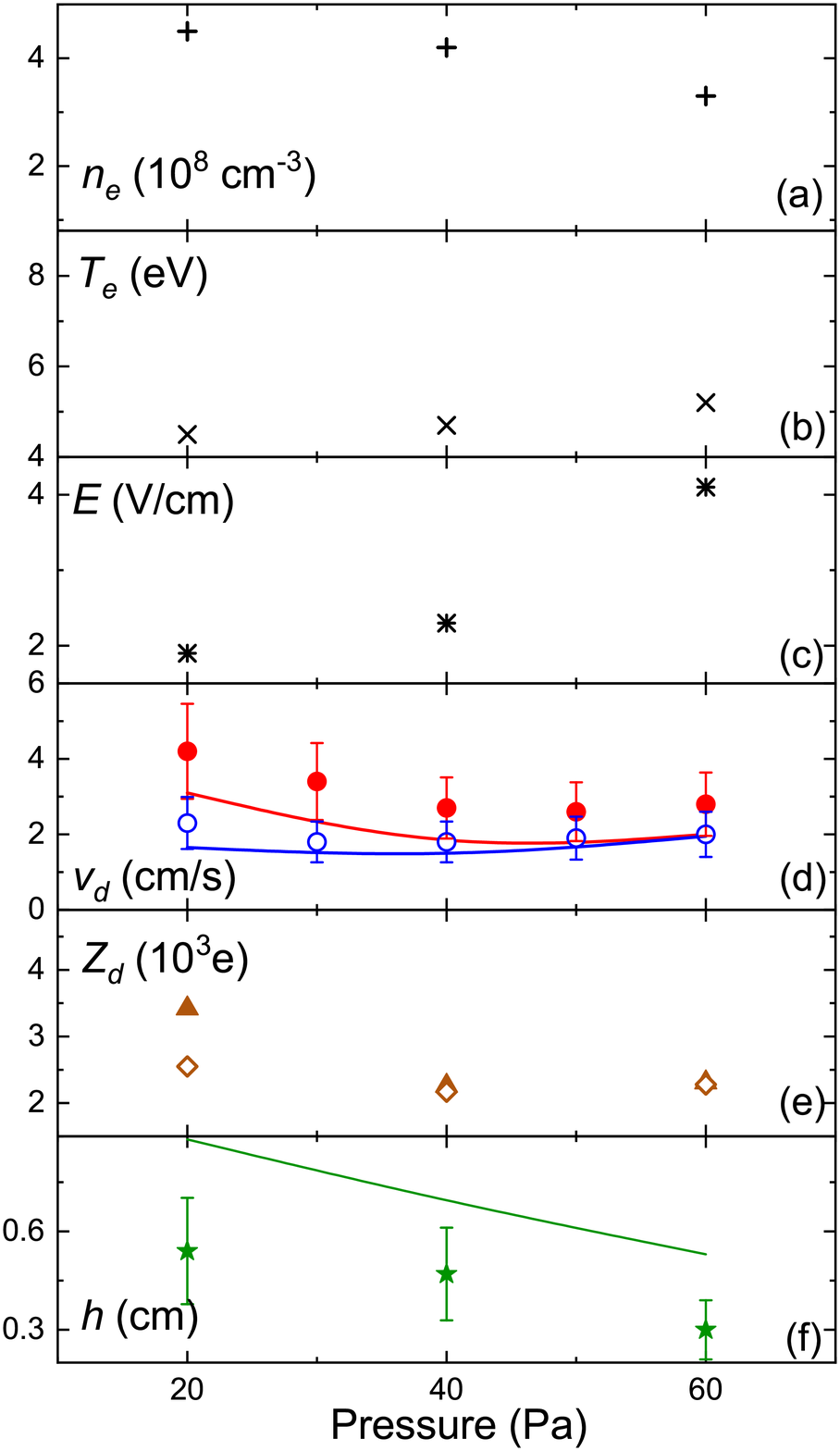}
	\caption{Same as in Fig. 3, but for argon gas. }
\end{figure}

These measurements showed that for neon plasma the electron density increases with pressure, the electron temperature slightly decreases and  the electric field remains almost the same in the chosen pressure range. At the same time, for argon plasma the measurements showed different dependence of the plasma parameters with pressure: the electron temperature and electric field increase with pressure.  The electron density shows the same dependence on the pressure as for neon.
These measured plasma parameters  or their interpolations (if the experimental conditions were different from the conditions in Ref.~\onlinecite{fortovmorfill}) were used in order to estimate the particle charge values in the dc discharge of PK-4.

\section{Theoretical model}

The theoretical model used in this work is based on the force balance condition. For simplicity, we consider a single particle and neglect possible effects of the particle component on the discharge parameters. As has been previously shown~\cite{yaroshenko}  even in a relatively dense cloud the electric force is dominant.  Under microgravity conditions, only the longitudinal direction is considered assuming that the test particle is located near the discharge axis. The main forces include the electric force, the ion drag force, and the neutral drag forces~\cite{fortov, fortovkhrapak} (the electron drag force can be also important for relatively low electron temperatures,~\cite{KhrapakPRE2004} but not for the conditions investigated here).  The microparticle drift velocity can be simply estimated from the force balance $F_{el}^{\parallel}+F_i^{\parallel}+F_n=0$, like in Ref.~\onlinecite{khrapak2013}. Here, $F_{el}^{\parallel}$ is a horizontal component of the electric force,  $F_i^{\parallel}$ is the horizontal component of the ion drag force and $F_n$ is the neutral drag force. In ground laboratory experiments, the particles levitating in the wall region experience a longitudinal as well as radial component of the electric  field and,  hence, the force balance includes here the radial component of the electric and ion drag forces as well as the gravity force $F_{el}^{\perp}+F_i^{\perp}+F_g=0$, were $F_g$ is the force of gravity.  The forces evaluation is supplemented by the appropriate charging model. The charge is calculated from the balance of electron and ion fluxes to the particle surface taking into account the effect of ion-neutral collisions~\cite{ zobnin2000, lampe, ratynskaia, khrapak2005,Khrapak2009} and ionization enhancement of the ion flux to the particle surface.~\cite{khrapak2012}      

The model used is very similar to that described in Ref.~\onlinecite{khrapak2013} which allows us to omit the details. The following improvements and updates have been implemented. 

The radial dependence of the plasma density and electric field has been explicitly introduced using the simplest approximation~\cite{reiser},    
\begin{equation}
n_i\simeq n_e \simeq n_0 J_0\left(2.4r/R\right),
\end{equation}
and
\begin{equation}
E_{\perp}=2.4\frac{T_e}{eR} \frac{J_1\left(2.4 r/R\right)}{ J_0\left(2.4 r/R\right)},
\end{equation}
where $n_0$ is the ion and electron density on the tube axis, $R$ is the tube radius, $T_e$ is the electron temperature, $e$ is the elementary charge, and $J_{0,1}(x)$ are Bessel functions of the first kind. Electron impact ionization rates were taken from Ref.~ \onlinecite{lennon}. Here, one needs to note that the  overestimation of the ionisation effect due to non-Maxwellian electron energy distribution  can lead to the underestimation of the microparticle  charge.~\cite{godyak}   

According to some recent results,~\cite{bronold, zobnin2018} the electron absorption coefficient for collisions with the particle surface was allowed to be different from unity. We have considered two cases, when it is equal to 1 and 0.5, respectively.  The estimations for these both cases do not differ much, therefore further we  present our results for the case  where the electron absorption coefficient is equal unity. Collisional enhancement of the ion drag force has been modeled using a correction factor proposed previously.~\cite{hutchinson}  We have also used a simple approximation for the effective ion-neutral collision frequency.~\cite{khrapakplasmaphys}

\section{Results} 

Figures 3 and 4 illustrate the comparison between experimental and theoretical results. 
Fig.~3 shows the dependence of  the measured plasma parameters (electron density  $n_e$, electron temperature $T_e$ and electric filed $E$)  on the gas pressure. In addition,  the experimental and theoretical drift velocity $v_d$, theoretically estimated microparticle charge $Z_d$ on ground  and under microgravity conditions and the distance of the upper microparticle layer from the center of the tube $h$ measured on ground and theoretically estimated are presented here. The data shown are for the 3.4 $\mu$m diameter particles in neon gas at 1 mA discharge current.  The experimental uncertainties are shown for the microparticle drift velocity $v_d$ and for the distance of the upper microparticle layer from the center of the tube $h$ and come from the determination of the microparticle positions. Here one can see that increasing the discharge pressure the microparticles become slower because of the neutral friction. Theoretically estimated microparticle drift velocity demonstrates that the chosen analytical model reproduces relatively well the experimental measurements.  

In addition,  the data of experiments performed on ISS confirm previous observations from parabolic flights,~\cite{khrapak2013} that microparticles  in ground-based experiments move faster than  under microgravity conditions. The difference is more pronounced for the lower pressure range (20 -- 30 Pa). The theoretical curves are in agreement with this tendency. The measured drift velocities in the case of neon gas for 3.4~$\mu$m diameter particles quantitatively agree  well with those measured during parabolic flights.~\cite{khrapak2013}     

Figure 4 shows similar dependencies for argon gas. 
Here, the microparticle velocity does not exhibit a monotonic decrease on the pressure as in neon case on Fig.~3. Instead, in the region of moderate pressures (60 Pa) the microparticle velocity somewhat increases with pressure. It can be explained by the nonmonotonic dependece of the plasma parameters with pressure, namely, electron temperature $T_{e}$ and electric field  $E$. Lower values of  $T_{e}$ in the case of argon compared to neon lead to a lower particle charge and, hence,  lower microparticle velocities. On the other hand, the electric field  $E$ is almost constant in the chosen pressure range.  The results from the theoretical model are consistent with this behavior.

The theoretically estimated charge of the 3.4~$\mu$m diameter particles presented for argon and neon in Fig.~3 and Fig.~4 shows also a different dependence on the pressure  for argon and neon gases. In the case of argon plasma the charge is estimated to be lower than in neon plasma. This is mainly related to the higher  electron temperature in neon gas.
 
 The model presented above includes the radial distribution of the discharge parameters within the discharge tube. This allows us to compare the distance of the upper microparticle layer from the center of the tube $h$ estimated from the experimental measurements on ground with that of a single particle obtained from theoretical calculations performed here. 
 
 The comparison in Fig.~3 and Fig.~4 for the case of neon and argon discharges shows some discrepancy between theory and experiments, although it shows better agreement in the case of neon gas.  The reason for this discrepancy could be the assumption of the single particle model in the theoretical consideration. Although in the experiments, we tried to keep the particle density as low as possible, we could not always achieve it. The estimations of the  microparticle influence on the electric field, as it has been discussed in Ref.~\onlinecite{usachev2016},  have showed that the electric field during the ISS experiments e.g. for the conditions presented in Fig.~2 (neon dc discharge, 60 Pa pressure and 1 mA discharge current) is about 30$\%$  higher than that measured by Langmuir probe in the particle-free discharge (see Fig.~3 and Fig.~4).

In addition, the number of particles could vary from one experiment to another. Finally, it cannot be excluded that the radial distribution of plasma density and electric field do not follow closely the simple theoretical approximation used in this work. However, a recent analysis of collective excitations~\cite{yaroshenko2019} for the wave experiments performed in PK-4, which includes the charge estimation from our model and assumes the so-called Havnes parameter approximately equal to unity, has not shown large inconsistency with the experiment.
 
Another possible reason for the deviation between the theory and experimental measurements could be the changing of the microparticle mass and shape during the exposure in the plasma discharge as it has been reported in several works.~\cite{pavlu2004,Carstensen2011,killer2016,Vysinka2018,karasev2019} Since in our case the time of the microparticles levitation in the discharge corresponds to a minute timescale, we do not expect this effect to be very important.  

\begin{figure}
	\center
	\includegraphics[width=7.5cm]{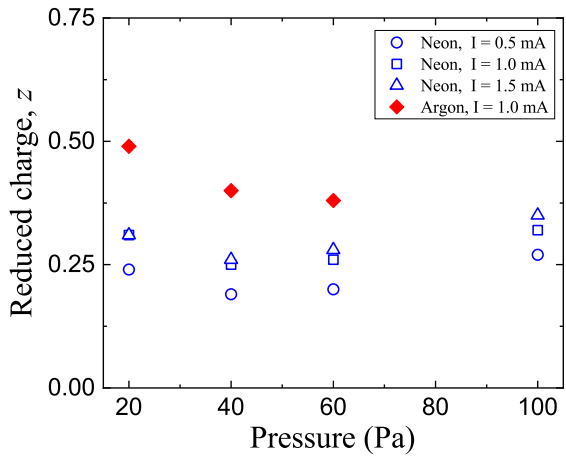}
	\caption{Reduced charge of 3.4 $\mu$m diameter particles at the discharge tube axis versus pressure  for neon and argon gases at several discharge current values. }
\end{figure} 

In Fig.~5 we have plotted the theoretically estimated reduced (normalized) charge $z=Z_{d}e^2/aT_e$, where $e$ is the elementary electron charge and $a$ is the radius of  the 3.4~$\mu$m diameter particles, versus pressure for argon plasma at $I=1.0$ mA and neon plasma at $I=0.5,1.0$ and $1.5$ mA. The charge calculation refers to the tube axis region. The reduced charge is essentially the particle surface potential in units of $T_{e}/e$.  The reduced charge is higher for argon plasma, mainly due to the lower electron-to-ion mass ratio.~\cite{khrapakpre} It does not vary much neither with pressure nor with discharge current in the investigated parameter regime. The averaged values are  $z$=0.3 $\pm $ 0.1 for neon and $z$=0.4 $\pm $ 0.1 for argon.  Since the reduced charge does not depend strongly on the particle size, the values shown should be generally representative for PK-4 experiments  under microgravity conditions.        

\section{Conclusion}

In this work the microparticle drift velocities from the analytical model, which takes into account the local distribution of the discharge parameters, were compared to those from the experiments performed  in a laboratory set-up on the ground and under microgravity conditions on the ISS. The experiments were carried out in a broad range of discharge parameters (pressure and discharge current) and in two different gases: argon and neon. Both the experimentally measured and theoretically estimated particle velocities and charges show different behavior on the pressure for argon and neon plasmas. Also the quantitative values of charge differ for these two gases. The reported values of particle charge can be used as reference values for other experiments with the PK-4 laboratory.  

\section{Acknowledgments}

All authors greatly acknowledge the joint ESA-Roscosmos “Experiment Plasmakristall-4” on board the International Space Station. This work was  partially supported by DLR Grants
Nos. 50WM1441 and 50WM1442. We also thank to Dr. Yaroshenko for comments and critical reading of the manuscript.

\bibliographystyle{unsrt}

\begin{thebibliography}{}

\end{thebibliography}


\begin{thebibliography}{}

\bibitem{fortov} V.~E.~Fortov, A.~V.~Ivlev, S.~A.~Khrapak, A.~G.~Khrapak, and G.~E.~Morfill, Phys. Rep. {\bf 421}, 1-103 (2005).

\bibitem{morfill} G.~E. Morfill, and A.~V.~Ivlev,  Rev. Mod. Phys. {\bf 81}, 1353-1404 (2009).

\bibitem{khrapak2013} S.~A. Khrapak, M.~H.~Thoma, M.~Chaudhuri, G.~E.~Morfill, A.~V.~Zobnin, A.~D.~Usachev,
O.~F.~Petrov, and V.~E.~Fortov, Phys. Rev. E {\bf 87}, 063109 (2013).

\bibitem{pustylnik} M.~Y.~Pustylnik, M.~A.~Fink, V.~Nosenko, T. Antonova,  T. Hagl, H. M. Thomas, A.~ V.~Zobnin, A .M. Lipaev, A. D. Usachev, V.I. Molotkov, {\it et al}., Rev. Sci. instrum. {\bf 87}, 093505 (2016).

\bibitem{weber} M.~Weber, M.~Fink, V.~Fortov, A.~Lipaev, V.~Molotkov, G.~Morfill, O.~Petrov, M.~Pustylnik, M.~Thoma, H.~Thomas, A.~Usachev, and C.~Raeth, Optics Express  {\bf 24}, 7987-8012 (2016).  

\bibitem{surabi} S. Jaiswal, M. Pustylnik,  S. Zhdanov, H. M. Thomas, A. M. Lipaev, A. D.Usachev, V. I. Molotkov, V. E. Fortov, M.H. Thoma, O.V. Novitskii, Phys. Plasmas  {\bf 25}, 083705 (2018).

\bibitem{yaroshenko2019} V. V. Yaroshenko,   S. A. Khrapak,  M. Y. Pustylnik, H. M. Thomas, S.~Jaiswal, A. M. Lipaev, A. D. Usachev,  O.F. Petrov, V. E. Fortov, Phys. Plasmas {\bf 26}, 053702  (2019).

\bibitem{fortovmorfill} V.~E.~Fortov, G.~E.~Morfill, O.~Petrov, M.~H.~Thoma, A.~D.~Usachev, H.~H\"{o}fner, A.~V.~Zobnin, M.~Kretschmer, S.~Ratynskaya, M.~A.~Fink, K.~Tarantik, Y.~Gerasimov, and V.~Esenkov, Plasma Phys. Controlled Fusion {\bf 47}, B537 (2005).

\bibitem{yaroshenko}  V.V. Yaroshenko, S.A. Khrapak and G. E. Morfill,  Phys. Plasmas {\bf 20}, 043703 (2013).

\bibitem{fortovkhrapak} V. E. Fortov, A. G. Khrapak, S. A. Khrapak, V. I. Molotkov, O. F. Petrov, Phys.-Usp. {\bf 47}, 447-492 (2004). 

\bibitem{KhrapakPRE2004} S. A. Khrapak and G. E. Morfill, Phys. Rev. E {\bf 69}, 066411 (2004).

\bibitem{zobnin2000}  A. V. Zobnin and A. P. Nefedov and V. A. Sinel'shchikov and V. E. Fortov, JETP {\bf 91}, 483--487 (2000). 

\bibitem{lampe} M. Lampe, R. Goswami, Z. Sternovsky, S. Robertson, V.Gavrishchaka, G. Ganguli and Glenn Joyce, Phys. Plasmas {\bf 10} 1500--1513 (2003). 

\bibitem{ratynskaia}  S. Ratynskaia, S. Khrapak,  A. Zobnin,  M. H. Thoma,  M. Kretschmer,  A. Usachev,  V. Yaroshenko, R. A. Quinn, G. E. Morfill,  O. Petrov and V. Fortov, Phys. Rev. Lett. {\bf 93}, 085001 (2004). 

\bibitem{khrapak2005} S. A. Khrapak, S. V. Ratynskaia,  A. V. Zobnin,  A. D. Usachev, V. V. Yaroshenko,  M. H. Thoma,  M. Kretschmer,  H. H\"{o}fner,  G. E. Morfill,  O. F. Petrov and V. E. Fortov, Phys. Rev. E {\bf 72}, 016406 (2005).

\bibitem{Khrapak2009} S. Khrapak and G. Morfill, Contrib. Plasma Phys. {\bf 49}, 148 (2009).

\bibitem{khrapak2012} S. A. Khrapak and G. E. Morfill, Phys. Plasmas {\bf 19}, 024510 (2012).

\bibitem{reiser} Yuri P. Raizer, Gas Discharge Physics  (Springer, 2011).

\bibitem{lennon} M. A. Lennon, K. L. Bell, H. B. Gilbody, J. G. Hughes, A. E. Kingston, M. J. Murray and F. J. Smith, J. Phys. Chem. Ref. Data {\bf 17}, 1285--1363 (1988).

\bibitem{godyak}  V. A. Godyak, R. B. Piejak and B. M. Alexandrovich Plasma Sources Sci. Technol. {\bf 11}, 525–543 (2002). 

\bibitem{bronold}  F. X. Bronold and H. Fehske, Phys. Rev. Lett. {\bf 115}, 225001 (2015).

\bibitem{zobnin2018} A.~V.~Zobnin, A.~D.~Usachev, O.~F.~Petrov,  V.~E.~Fortov,  M.~H.~Thoma, and  M.~A.~Fink, Phys. Plasmas  {\bf 25}, 033702 (2018).

\bibitem{hutchinson} I. H. Hutchinson and C. B. Haakonsen, Phys. Plasmas {\bf 20}, 083701 (2013).

\bibitem{khrapakplasmaphys}  S. Khrapak, J. Plasma Phys. {\bf 79} 1123--1124 (2013). 

\bibitem{usachev2016}  A. D. Usachev, A. V. Zobnin, O. F. Petrov, V. E. Fortov, M. H. Thoma,
M. Y. Pustylnik, M. A. Fink and G. E. Morfill, Plasma Sources Sci. Technol. {\bf 25}, 035009 (2016).

\bibitem{pavlu2004} Ji\v{r}\'{i} Pavl\r{u}, Andriy Velyhan, Ivana Richterov\'{a}, Zden\v{e}k N\v{e}me\v{c}ek, Jana  \v{S}afr\'{a}nkov\'{a}, Ivo \v{C}erm\'{a}k, and Peter \v{Z}ilav\'{y}, IEEE Trancactions on Plasma Science {\bf 32}, 2 (2004).

\bibitem{Carstensen2011} J. Carstensen, H. Jung, F. Greiner, and A. Piel, Phys. Plasmas {\bf 18}, 033701 (2011).

\bibitem{killer2016} C. Killer, F. Greiner, S. Groth, B. Tadsen
and A. Melzer,  Plasma Sources Sci. Technol. {\bf 25},  055004 (2016). 

\bibitem{Vysinka2018} M. Vysinka, L. Nouzak, J. Pavlu, Z. Nemecek, J. Safrankova, I. Richterova, IEEETrans. on Plasma Sci. 46, 709 (2018).

\bibitem{karasev2019} V. Yu. Karasev,  E. S. Dzlieva, S. I. Pavlov, O. V. Matvievskaya,
V. A. Polischuk, M. A. Ermolenko, A. I. Eikhval’d, A. P. Gorbenko, Contributions to Plasma Physics e201800145 (2019) https://doi.org/10.1002/ctpp.201800145

\bibitem{khrapakpre}  S. A. Khrapak, A. V. Ivlev, and G. Morfill,  Phys. Rev. E {\bf 64}, 046403  (2001).

\end{thebibliography}

\end{document}